\documentclass[aps,amsfonts,prl,nofootinbib,tightenlines,twocolumn,showpacs]{revtex4}

\usepackage{bm}
\usepackage{epsfig}

\newcommand{\sech}{{\, \rm sech \,}}
\newcommand{\arxiv}[1]{{#1}}
\newcommand{\currtime}{}

\begin{document}

\title{An Electroweak Oscillon}

\author{N.\ Graham}
\email{ngraham@middlebury.edu}
\affiliation{Department of Physics, Middlebury College,
Middlebury, VT  05753} 
\affiliation{Center for Theoretical Physics, Laboratory for
Nuclear Science, and Department of Physics,
Massachusetts Institute of Technology,
Cambridge, MA 02139}

\preprint{\rm MIT-CTP\# 3777 \qquad NSF-KITP-06-68 \qquad hep-th/yymmnnn}

\begin{abstract}

A recent study demonstrated the existence of oscillons --- extremely
long-lived localized configurations that undergo regular oscillations in
time --- in spontaneously broken $SU(2)$ gauge theory with a fundamental
Higgs particle whose mass is twice the mass of the gauge bosons.  This
analysis was carried out in a spherically symmetric ansatz
invariant under combined spatial and isospin rotations.   We extend
this result by considering a numerical simulation of the the full
bosonic sector of the $SU(2)\times U(1)$ electroweak Standard Model in
$3+1$ dimensions, with no assumption of rotational symmetry, for a
Higgs mass equal to twice the $W^\pm$ boson mass.  Within the limits of
this numerical simulation, we find that the oscillon solution from the
pure $SU(2)$ theory is modified but remains stable in the full
electroweak theory.  The observed oscillon solution contains total
energy approximately 30\ TeV localized in a region of radius
approximately 0.05\ fm.
 
\end{abstract}

\maketitle

\paragraph{Introduction}

In nonlinear field theories, static soliton solutions to the equations
of motion have been well studied (see for example
\cite{Coleman,Rajaraman}).  However, a much broader class of theories
contain oscillon solutions, which are localized in space but oscillate
in time.  In some special cases, such as the sine-Gordon breather
\cite{DHN} and $Q$-ball \cite{ColemanQ}, conserved charges guarantee
the existence of exact, periodic solutions.  Even in the absence of
such guarantees, however, localized solutions have been found in many
different theories that either live indefinitely or for extremely long
times compared to the natural timescales of the system.

For scalar theories in one space dimension, oscillon solutions have
been found to remain periodic to all orders in a perturbative
expansion \cite{DHN} and are never seen to decay in numerical
simulations \cite{Campbell}, but can decay after extremely long times
via nonperturbative effects \cite{Kruskal} or by coupling to an
expanding background \cite{oscex}.  Both $\phi^4$ theory in two
dimensions \cite{2d1,2d2} and the abelian Higgs model in one dimension
\cite{abelianhiggs} have also been shown to contain oscillon solutions
that are not observed to decay.  In $\phi^4$ theory in three dimensions
\cite{Bogolyubsky,Gleiser,Honda,iball,Forgacs}, however, there exist
long-lived quasi-periodic solutions whose lifetime depends sensitively on
the initial conditions.  Similar behavior is present in other scalar
theories in three dimensions \cite{Wojtek} and in higher dimensions
\cite{Gleiserd}.  Phenomenologically, small $Q$-balls were considered as
dark matter candidates in \cite{smallq1,smallq2,Enqvist,Kasuya}, axion
oscillons were considered in \cite{Kolb}, and the possible role of
oscillons in and after inflation was studied in
\cite{McDonald,Rajantie,Gleiserinflat}. Oscillon-like solutions have
also been studied in connection with phase transitions
\cite{Gleiserphase}, monopole systems \cite{monopole}, QCD
\cite{Hsu}, and gravitational systems \cite{Khlopov}.

A recent paper \cite{oscillon} demonstrated numerical evidence for an
oscillon solution in spontaneously broken $SU(2)$ gauge theory with a
fundamental Higgs whose mass is exactly twice that of the gauge bosons.
Current work \cite{twofield} is investigating an
analytic explanation of this mass relationship using a small amplitude
analysis \cite{DHN,Hsu,smallamp}, in which the $2:1$
ratio arises as a resonance condition necessary for quadratic
nonlinear terms to balance dispersive linear terms in the equations of
motion.  In this analysis, a field of mass $m$ oscillates with
amplitude $\epsilon m$, frequency $m\sqrt{1-\epsilon^2}$, and length
scale $1/(\epsilon m)$.  A similar mass relation arises in the study of
embedded defects \cite{Lepora}.  The field configurations in
\cite{oscillon} were restricted to the spherical ansatz
\cite{spherical}, meaning they were invariant under combined rotations
in space and isospin.  Here we extend this analysis to a fully
three-dimensional spatial lattice, eliminating any symmetry
assumptions.  We include the $U(1)$ hypercharge field, so that we are
simulating the full electroweak sector of the Standard Model without
fermions.  We use the same $SU(2)$ gauge coupling $g$ and Higgs
self-coupling $\lambda$ as in the pure $SU(2)$ theory, meaning that
the Higgs mass is twice the mass of the $W^{\pm}$ bosons, and set the
$U(1)$ coupling $g'$ so that the mass of the $Z^0$ boson matches its
observed value.

While one might expect the oscillon to decay rapidly by emitting
electromagnetic radiation, it does not.  Instead, after initially
shedding some energy into electromagnetic radiation, the system
settles into a stable, localized oscillon solution that no longer
radiates.  Similar behavior was observed both when an additional
massless scalar field was coupled to breathers in one-dimensional
$\phi^4$ theory and when an additional spherically symmetric massless
scalar field was coupled to oscillons in the spherical ansatz, results
that provided motivation for this work.

\paragraph{Continuum Theory}

We begin from $SU(2)\times U(1)$ electroweak theory in the continuum,
ignoring fermions.  We follow the standard classical treatment of
spontaneously broken nonabelian field theory (see for example
\cite{Huang}).  The Lagrangian density is 
\begin{equation}
{\cal L} = (D_\mu \Phi)^\dagger D^\mu \Phi -\frac{1}{4} F_{\mu \nu}
F^{\mu \nu} -\frac{1}{4} {\bm F}_{\mu \nu} \cdot {\bm F}^{\mu \nu}
- \lambda(|\Phi|^2 - v^2)^2 \,,
\end{equation}
where the boldface vector notation refers to isovectors.  Here $\Phi$
is the Higgs field, a Lorentz scalar carrying $U(1)$
hypercharge $1/2$ and transforming under the fundamental
representation of $SU(2)$.  The metric signature is $+---$.
The $SU(2)$ and $U(1)$ field strengths are
\begin{eqnarray}
\bm{F}_{\mu \nu} &=& \partial_\mu {\bm W}_\nu - \partial_\nu \bm{W}_\mu
- g{\bm W}_\mu\times {\bm W}_\nu \,, \cr
F_{\mu \nu} &=& \partial_\mu B_\nu - \partial_\nu B_\mu \,,
\end{eqnarray}
and the covariant derivatives are given by
\begin{eqnarray}
D_\mu \Phi &=& \left(\partial_\mu + i \frac{g'}{2} B_\mu
+ i \frac{g}{2} \bm{\tau} \cdot \bm{W}_\mu\right)\Phi \,, \cr
D^\mu \bm{F}_{\mu \nu} &=& \partial^\mu \bm{F}_{\mu \nu}
- g \bm{W}^\mu \times \bm{F}_{\mu\nu} \,,
\end{eqnarray}
where $\bm{\tau}$ represents the weak isospin Pauli matrices.
We obtain the equations of motion
\begin{eqnarray}
\partial_\mu F^{\mu\nu} &=& J^\nu \,, \cr
D_\mu \bm{F}^{\mu\nu} &=& \bm{J}^\nu \,, \cr
D^\mu D_\mu \Phi &=& 2\lambda(v^2 - |\Phi|^2) \Phi\,,
\end{eqnarray}
where the gauge currents are
\begin{eqnarray}
J_\nu &=& g' {\, \rm Im \,} (D_\nu \Phi)^\dagger \Phi\,, \cr
\bm{J}_\nu &=& g {\, \rm Im \,} (D_\nu \Phi)^\dagger \bm{\tau} \Phi \,.
\end{eqnarray}

We work in the gauge $B_0 = 0$, $\bm{W}_0 = \bm{0}$.  With this choice, the
covariant time derivatives become ordinary derivatives and we can
apply a Hamiltonian formalism.  The energy density is
\begin{eqnarray}
u &=& \frac{1}{2}\sum_{j=x,y,z}\Bigg[
\dot B_j^2 + \bm{\dot W}_j\cdot\bm{\dot W}_j 
+ \sum_{k>j} \left(
F_{kj}^2 + \bm{F}_{kj} \cdot \bm{F}_{kj} \right) \Bigg]
\cr &+&
|\dot \Phi|^2 + \sum_{j=x,y,z} (D_j \Phi)^\dagger (D_j \Phi)
+ \lambda\left(|\Phi|^2 - v^2\right)^2 \,,
\end{eqnarray}
whose integral over space is conserved.  Here dot indicates time
derivative.  From the equations for $B_0$ and $\bm{W}_0$, we obtain
the Gauss's Law constraints,
\begin{eqnarray}
\sum_{j=x,y,z} \partial_j \dot B_j - J_0 &= 0\,, \cr
\sum_{j=x,y,z} D_j \bm{\dot W}_j - \bm{J}_0 &=& 0\,,
\end{eqnarray}
where the charge densities are
\begin{eqnarray}
J_0 &=& g' {\, \rm Im \,} \dot \Phi^\dagger \Phi \,, \cr
\bm{J}_0 &=& g {\, \rm Im \,} \dot \Phi^\dagger \bm{\tau}\Phi \,.
\end{eqnarray}
These constraints remain true at all times, at all points in space,
assuming they are obeyed by the initial value data.

\paragraph{Lattice Theory}

We use the standard Wilsonian approach \cite{Wilson} for implementing gauge
fields on the lattice (for a review see \cite{SmitBook}), adapted to
Minkowski space evolution as in \cite{Shaposhnikov,SmitSim,RajantieSim}. 
(The details of the discretization have been modified slightly for the
present application.)  The $U(1)$ and $SU(2)$ gauge fields live on the
links of the lattice and the fundamental Higgs field lives at the lattice
sites.  The lattice spacing is $\Delta x$, and we determine the values of
the fields at time $t_+ = t+\Delta t$ based on their values at times $t$
and $t_- = t-\Delta t$.  We associate with the link emanating from lattice
site $p$ in the positive $j^{\rm th}$ direction the Wilson line 
\begin{equation}
U_j^p = e^{i g' B_j^p \Delta x /2}  
e^{i g \bm{W}_j^p\cdot\bm{\tau} \Delta x/2} \,.
\end{equation}
We define the Wilson line for the link emanating from lattice site $p$
in the negative $j^{\rm th}$ direction to be the adjoint of the
corresponding Wilson line emanating in the positive direction from the
neighboring site, $U_{-j}^p = (U_j^{p-j})^\dagger$,
where the notation $p \pm j$ indicates the adjacent lattice site to
$p$, displaced from $p$ in direction $\pm j$.  At the edges of the
lattice we use periodic boundary conditions.  We define the logarithm
of $2\times 2$ matrices of this form as
\begin{equation}
\ln U_j^p = \frac{i\Delta x}{2} (g' B_j^p +  g
\bm{W}_j^p\cdot\bm{\tau}) \,,
\end{equation}
and note that $\ln AB \neq \ln A + \ln B$ when the matrices do not commute.

The equation of motion for the Higgs field at site $p$ is
\begin{equation}
\Phi^p(t_+) = 2 \Phi^p\currtime - \Phi^p(t_-) + \Delta t^2 \ddot
\Phi^p \,,
\end{equation}
where
\begin{equation}
\ddot \Phi^p = \sum_{j=\pm x,\pm y, \pm z}
\frac{U_j^p\currtime \Phi^{p+j}\currtime - \Phi^p\currtime}{\Delta
x^2} + 2\lambda\left(v^2 - |\Phi^p\currtime|^2\right)\Phi^p\currtime \,,
\end{equation}
and all fields are evaluated at time $t$ unless otherwise indicated.
For the gauge fields, we have
\begin{eqnarray}
U_j^p(t_+) =
\exp\Bigg[\ln U_j^p\currtime U_j^p(t_-)^\dagger
+\Bigg( \frac{\Delta x}{2i} (
g' J_j^p + g \bm{J}_j^p \cdot \bm{\tau} ) \cr
- \sum_{j'\neq j} 
\frac{\ln U^p_{\square(j,j')}\currtime + \ln U^p_{\square(j,-j')}\currtime
}{\Delta x^2}\Bigg)\Delta t^2 \Bigg] U_j^p\currtime \,,~~~
\end{eqnarray}
where $U^p_{\square(j,j')}\currtime = U_j^p\currtime
U_{j'}^{p+j}\currtime U_{-j}^{p+j+j'}\currtime
U_{-j'}^{p+j'}\currtime$ and
\begin{eqnarray}
J_j^p &=& g' {\rm \, Im \,}
\frac{(\Phi^p\currtime)^\dagger U_{j}^p\currtime
\Phi^{p+j}\currtime}{\Delta x}\,, \cr
\bm{J}_j^p &=& g {\rm \, Im \,} 
\frac{(\Phi^p\currtime)^\dagger  \bm{\tau} U_{j}^p\currtime
\Phi^{p+j}\currtime}{\Delta x} \,,
\end{eqnarray}
are the gauge currents.  The energy density at $p$ is then
\begin{eqnarray}
u^p\currtime &=& 
\sum_{j=x, y, z} \Bigg[ 
\frac{1}{2} \frac{\left\|\exp \left(\ln U_j^p(t_+) -
\ln U_j^p(t_-) \right)\right\|^2}{(2 \Delta t)^2} \cr
&+& \frac{1}{2} \sum_{j'> j}
\frac{\left\| U^p_{\square(j,j')}\currtime\right\|^2}{\Delta x^2}
+ \frac{\left|U_j^p\currtime\Phi^{p+j}\currtime 
- \Phi^{p}\currtime \right|^2}{\Delta x^2} \Bigg] \cr
&+& \frac{\left|\Phi^{p}(t_+) - \Phi^{p}(t_-) \right|^2} {4\Delta t^2}
+ \lambda\left(|\Phi^p|^2 - v^2\right)^2 \,,
\end{eqnarray}
whose integral over the whole lattice is conserved.  Here we have
defined 
\begin{eqnarray}
\left\|U_j^p\right\|^2 &=& 
\frac{\left|{\rm Tr\,} \ln U_j^p\right|^2}{g'^2 \Delta x^2} +
\frac{\left({\rm Tr\,} \bm{\tau} \ln U_j^p \right)^\dagger \cdot \left(
{\rm Tr\,} \bm{\tau} \ln U_j^p \right)}{g^2 \Delta x^2} \cr
&=& |B_j^p|^2 + \bm{W}_j^p\cdot\bm{W}_j^p
\end{eqnarray}
for any $U(2)$ link matrix.

At every lattice point, Gauss's Law,
\begin{eqnarray}
\sum_{j=x, y, z}
\frac{\ln U_j^p(t_+)    U_j^p(t_-)^\dagger
 +    \ln U_{-j}^p(t_+) U_{-j}^p(t_-)^\dagger}
{i \Delta x^2 \Delta t} \cr
- \left(g' J_0^p + g \bm{J}_0^p \cdot \bm{\tau}\right) = 0\,,
\label{Gauss}
\end{eqnarray}
is also maintained throughout the evolution, where the charge
densities are given by
\begin{eqnarray}
J_0 &=& g' {\rm \, Im \,} \left(\frac{\Phi^{p}(t_+) -
\Phi^{p}(t_-)} {2 \Delta t} \right)^\dagger \Phi^{p}\currtime \,,
\cr
\bm{J}_0 &=& g {\rm \, Im \,} \left(\frac{\Phi^{p}(t_+) -
\Phi^{p}(t_-)} {2 \Delta t} \right)^\dagger \bm{\tau}
\Phi^{p}\currtime \,.
\end{eqnarray}

\paragraph{Numerical Simulation}

The initial conditions for the simulation are obtained starting from
an approximate functional fit to the oscillon solutions found in 
numerical simulations of the $SU(2)$-Higgs theory in the
spherical ansatz \cite{oscillon}.  This result provides the
initial data for the $\bm{W}$ and $\Phi$ fields, and the initial $B$
field is chosen to vanish.  In order to guarantee that
the initial configuration continues to obey Gauss's Law in the full
$SU(2)\times U(1)$ theory, we generate the fit at a point in the cycle
where the time derivatives are smallest, and impose the requirement that
all time derivatives vanish for our initial data.\footnote{Even for
pure $SU(2)$ theory, an approximate fit with nonvanishing time
derivatives will no longer obey Gauss's Law.  In that case, however,
one can restore Gauss's Law by adjusting $\Phi(t_+)$ slightly via an
$SU(2)$ transformation at each point,
\begin{equation}
\Phi_{\rm new}(t_+) = \left|\frac{\Phi_{\rm old}(t_+)}{\Phi\currtime}\right|
{\cal U}^p \Phi\currtime
\end{equation}
with
\begin{equation}
{\cal U}^p = \exp\left[
\sum_{j=x, y, z}
\frac{\ln U_j^p(t_+) (U_j^p\currtime)^\dagger
 + \ln U_{-j}^p(t_+) (U_{-j}^p\currtime)^\dagger }
{g^2 \Delta x^2   |\Phi_{\rm old} (t_+)| |\Phi\currtime|/2}
\right]^\dagger \,.
\end{equation}
In the $SU(2) \times U(1)$ theory, this procedure is no longer
possible because $\Phi$ carries both charges, so it would be necessary
to adjust the $U(1)$ field as well, which cannot be done locally.}

The initial conditions are of the spherical ansatz form
\begin{eqnarray}
\bm{\tau} \cdot 
\bm{W}_i &=& 
\frac{1}{g}\left[
a_1(r,t) \bm{\tau}\cdot \bm{\hat x} \hat x_i +
\frac{\alpha(r,t)}{r}(\tau_i - {\bm{\tau}}\cdot\bm{\hat x}\hat x_i)
\right. \cr && \left.
 - \frac{\gamma(r,t)}{r}(\bm{\hat x} \times \bm{\tau})_i \right] ,
\cr
\Phi &=& \frac{1}{g} \left[\mu(r,t) - i \nu(r,t) \,
{\bm{\tau}}\cdot\bm{\hat x}\right]
\pmatrix{0 \cr 1} \,,
\label{ansatz}
\end{eqnarray}
where $\bm{\hat x} = \bm{x}/r$ and $r$ is the distance from the origin. The
field definitions have been chosen so that the reduced fields match those
used in \cite{oscillon}, even though the conventions for the
three-dimensional theory used here are slightly different.  We work in
units where $v=1/\sqrt{2}$.  Since we are dealing with classical dynamics,
we can rescale the fields to fix the $SU(2)$ coupling constant at $g =
\sqrt{2}$, so that the $W^\pm$ mass is then $m_W=gv/\sqrt{2}=1/\sqrt{2}$. 
With this rescaling, we must also introduce an overall factor of
$g^2/g_W^2$ multiplying the total energy, where $g_W=0.634$ is the true
weak coupling constant.  We choose $\lambda=1$, so that the Higgs mass is
twice the $W^\pm$ mass, $m_H=2v\sqrt{\lambda}=\sqrt{2}$.  These choices of
units and normalization agree with \cite{oscillon}, except the quantity
called $v$ here is smaller than $v$ in \cite{oscillon} by a factor of
$\sqrt{2}$. Finally, we fix $g'=0.773$, so that the ratio $g'/g$ matches
the observed value and the $Z^0$ boson has the correct mass.

In these units, we consider initial configurations
\begin{eqnarray}
a_1(r) &=& \epsilon (0.117 \epsilon + 0.016 \epsilon r)
\left(\sech 2 \epsilon r\right)^{1/8} \,, \cr
\mu(r) &=& 1 - 0.138 \epsilon \sech \frac{\epsilon r}{6.75}\,, \cr
\nu(r) &=& 0.017 \epsilon r \sech \frac{\epsilon r}{5}\,, \cr
\alpha(r) &=& 0.117 \epsilon^2 r \sech\frac{\epsilon r}{8}\,, \cr
\gamma(r) &=& 0 \,,
\label{initial}
\end{eqnarray}
where the adjustable parameter $\epsilon$ allows us to include a
combined rescaling of the fields' amplitudes and $r$-dependence, as is
commonly used in a small amplitude analysis \cite{DHN,Hsu,smallamp}.
While $\epsilon=1$ gives an approximation to the spherical ansatz
solution of \cite{oscillon}, a slightly larger value appears to be
necessary for the configuration to settle into a stable solution in
the full $SU(2)\times U(1)$ model.  The first term in parentheses in
the definition of $a_1(r)$ is scaled with an additional $\epsilon$ so
that it matches the coefficient of $\alpha$, ensuring that $\alpha$,
$a_1-\alpha/r$, $\gamma/r$, and $\nu$ all vanish as $r \to 0$, as
required for regularity of the fields at the origin.  Within the
spherical ansatz simulation, these initial conditions settle into a
long-lived oscillon in the pure $SU(2)$-Higgs theory, which we never
see decay.  The $U(1)$ interactions break the grand spin symmetry of
the spherical ansatz, though the continuum theory still preserves
invariance under combined space and isospin rotations around the
$z$-axis.  The Cartesian lattice provides a small breaking of all
rotational symmetries.  Thus configurations at later times are not
constrained to lie within this reduced ansatz.  (The full
three-dimensional simulation does continue to agree with the spherical
ansatz simulation for the case when the $U(1)$ interaction is turned
off, another check of the numerical calculation.)

We start from these initial conditions and let the system evolve for
as long as is practical numerically.  One concern is that
the outgoing radiation emitted as the configuration settles
into the oscillon solution can wrap around the periodic boundary
conditions, return to the region of the oscillon, and potentially
destabilize it.  However, as long as the region in which the oscillon
is localized does not represent a significant fraction of the lattice
volume, this radiation is sufficiently diffuse that it does not affect
the oscillon.  We use a lattice of size $L=144$ on a side in natural
units, which is more than enough to satisfy this criterion.
For $L \gtrsim 100$, changing the lattice size simply changes the
pattern of noise caused by electromagnetic radiation superimposed on
the oscillon region, but does not affect oscillon properties or
stability.  We can therefore be certain that there is no coherent
structure to this unphysical radiation that could possibly be
necessary for the oscillon's stability; its only possible effect is to
destabilize the oscillon, which only occurs if the radiation is
artificially concentrated by a small lattice (e.g. of size $L < 100$).
In numerical experiments, these destabilization effects are actually
much weaker in the electroweak model than in pure scalar or $SU(2)$
Higgs-gauge models, because in the electroweak model the radiated
energy is almost entirely in the electromagnetic field, while the
oscillon solution arranges itself to be electrically neutral.
For this reason, it is not necessary to use absorptive techniques such
as adiabatic damping \cite{2d1} (which would have to be adapted to
accommodate gauge invariance) or an expanding background \cite{oscex}.

We use lattice spacing $\Delta x = 0.75$, though $\Delta x = 0.625$
and $\Delta x = 0.25$ were verified to give completely equivalent
results in smaller tests.  The time step is $\Delta t=0.1$.  Total
energy is conserved to a few parts in $10^3$, which is appropriate
since our algorithm is second-order accurate.  To check Gauss's Law,
we square the left-hand side of Eq.\ (\ref{Gauss}), take the
trace, and then take the square root of the result.  The integral of
this quantity over the lattice never exceeds $0.025$ and shows no
upward trend over time, a highly nontrivial check on the numerical
calculation.  It is necessary, however, to use double precision to
avoid very gradual degradation in this result.\footnote{Here we have
computed the Gauss constraint at time $t$, which is obeyed to order
$\Delta t^2$.  The Gauss constraint at time $t+\Delta t/2$ is obeyed
to machine precision throughout the numerical evolution.}  A run to
time $10,000$ takes roughly $40$ hours using $24$ parallel processes,
each running on a $2\ {\rm GHz}$ Opteron processor core.

\begin{figure}[htbp]
\includegraphics[width=0.95\linewidth]{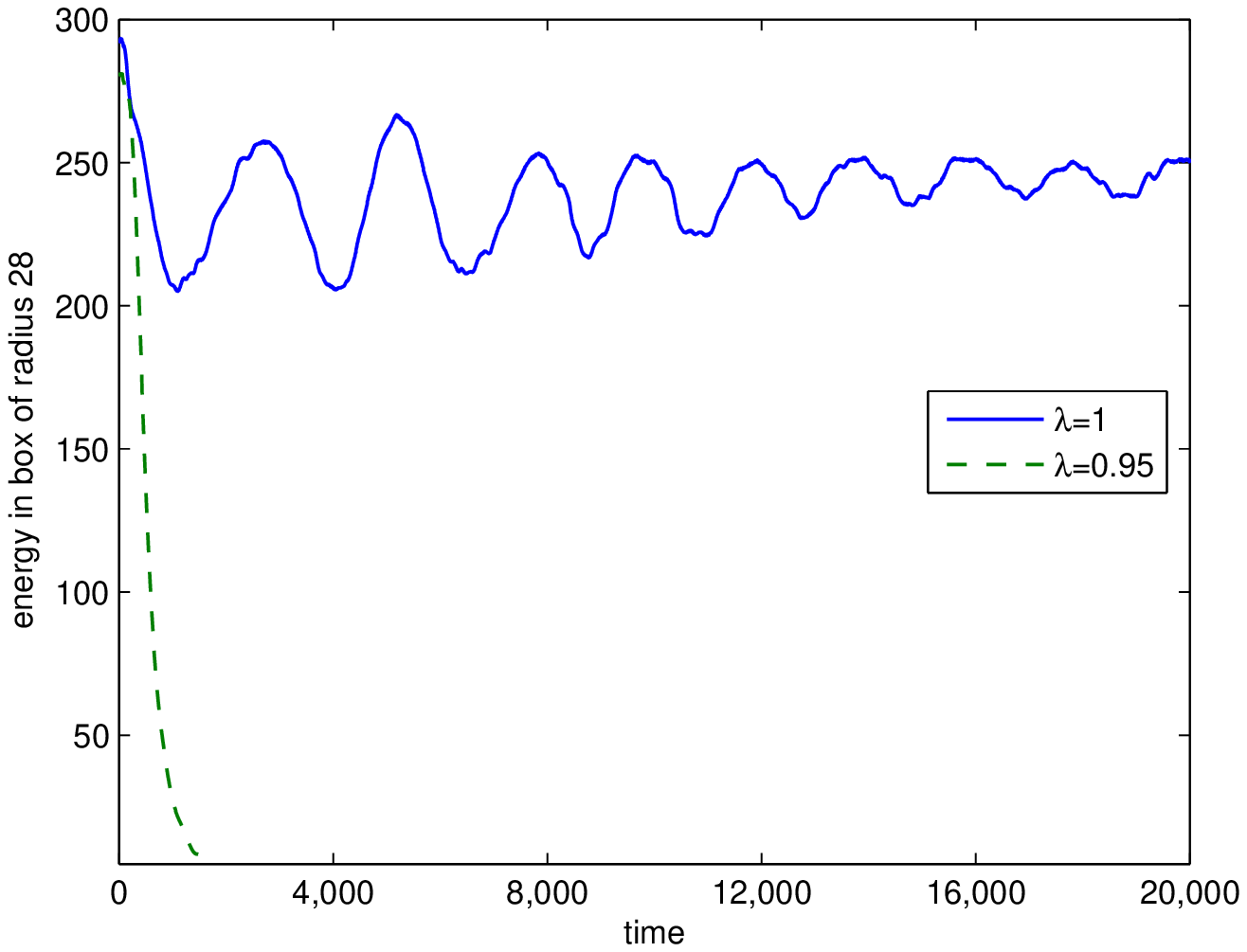}
\includegraphics[width=0.95\linewidth]{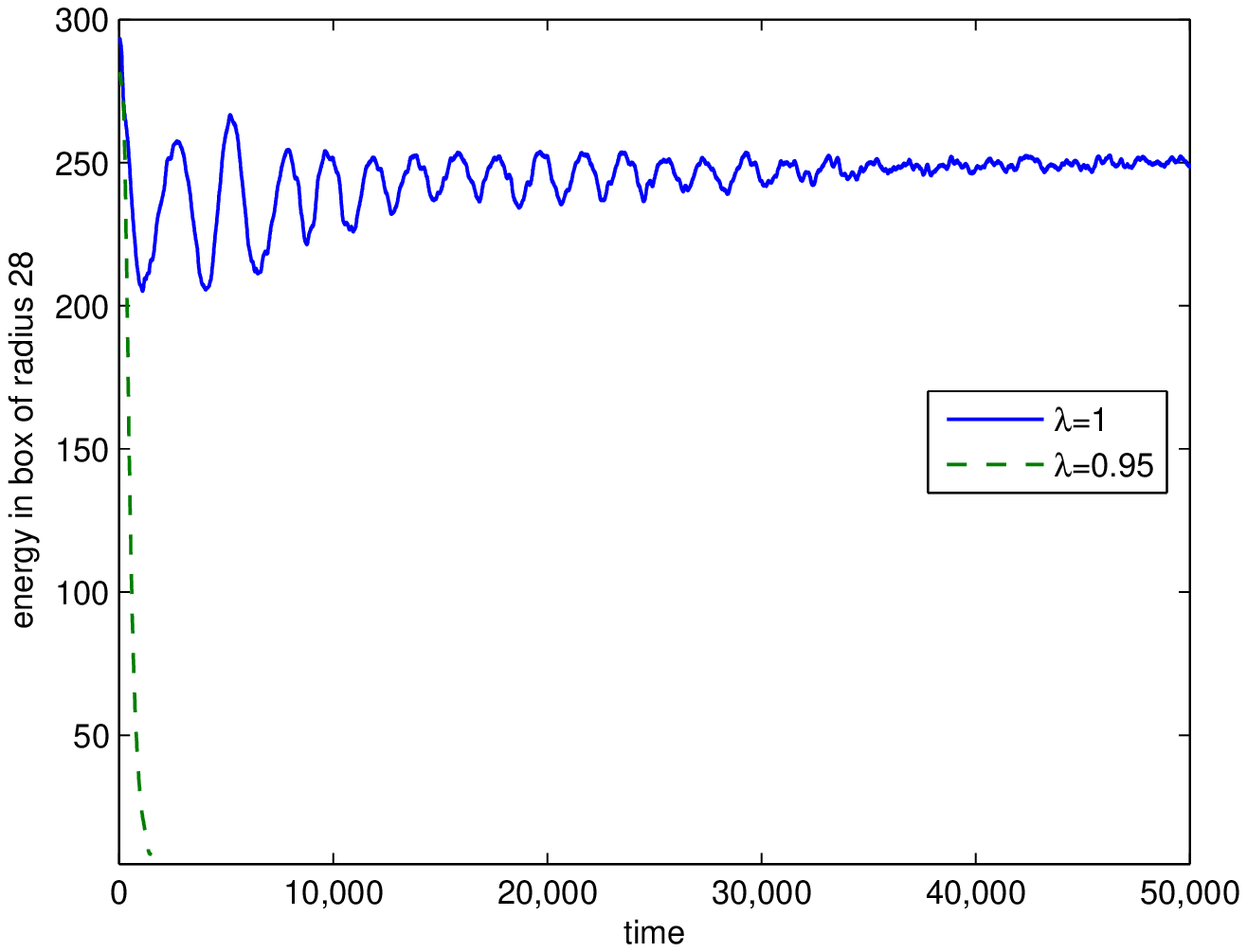}
\caption{Energy in a box of radius 28 as a function of
time in the natural units of \cite{oscillon}.
One unit of energy is $114\ {\rm GeV}$, one unit of time is
$5.79 \times 10^{-27}\ {\rm sec}$, and one unit of length is
$1.74 \times 10^{-18}\ {\rm m}$.
In the top panel, the initial conditions are given by Eq.\
(\ref{initial}) with $\epsilon=1.15$.  In the bottom panel, the
initial conditions are the same except the $\tau_z$ component of
$\bm{W}$ in Eq.\ (\ref{ansatz}) has been set to zero.  Two values of
the Higgs self-coupling $\lambda$ are shown.  For $\lambda=1$, the
masses of the Higgs and $W$ fields are in the $2:1$ ratio needed for
oscillon formation and the solution remains localized throughout the
simulation.  A transient beat pattern is also visible.  For
$\lambda=0.95$, the mass ratio is $1.95:1$.  In that case, there is no
stable object and the solution quickly disperses.}
\label{fig}
\end{figure}

Figure\ \ref{fig} shows the energy in a spherical box of
radius $28$ as the fields are evolved from the initial conditions in
Eq.\ (\ref{initial}).  When the Higgs mass is twice the $W^\pm$ mass, 
only a small amount of energy is emitted from the central region, with
the rest remaining localized for the length of the simulation in a
stable oscillon solution.  If the masses are not in this ratio,
however, the initial configuration quickly disperses.   The box radius
has been chosen to be just large enough to enclose essentially all of
energy density associated with the stable oscillon solution.  As a
result, as the initial conditions settle into the stable oscillon
solution, we are also able to see a transient ``beat'' pattern:  the
field configurations gradually expand and contract slightly
over many periods, causing a small amount of energy to move in and out
of the box, accompanied by a corresponding modulation of the field
amplitudes.  (When a larger box size is used, the graph of the energy in
the box flattens out.)  Similar beats appear in the $SU(2)$ spherical
ansatz oscillon \cite{oscillon}, but in the electroweak oscillon their
amplitude decays more rapidly.  As in the spherical ansatz oscillon, 
in the electroweak oscillon each excited field oscillates at a
frequency just below its mass, with amplitude of order $0.1$ and
typical radius of order $10$ in our units.  By comparing the total
number of cycles to the total time, we find $\omega_H = 1.404$ for
the Higgs field components and $\omega_W=0.702$ for the gauge field
components.  The primary excitations are in the $W^\pm$ fields and the
$\Phi$ field, with some energy radiated outward in the electromagnetic
field in a dipole pattern and the $Z^0$ field largely absent.  In
contrast, in the spherical ansatz oscillon the $W^\pm$ and $Z^0$
fields must appear symmetrically.  The electroweak oscillon does
remain approximately axially symmetric under combined space and
isospin rotations around the $z$-axis.  These results suggest a simple
modification of the initial configurations in which the $\tau_z$
component of $\bm{W}$ is set to zero in Eq.\ (\ref{ansatz}).
This modification yields an equivalent final oscillon configuration,
with the same field amplitudes, localized energy and field
frequencies.  However, less energy is shed initially, so there is less
superimposed noise caused by radiation returning from the boundaries.
As a result, the beat pattern is also more clearly visible.  This case
is also shown in Fig.\ \ref{fig}.

\paragraph{Conclusions}

We have seen strong evidence for the existence of a long-lived, localized,
oscillatory solution to the field equations of the bosonic electroweak
sector of the Standard Model in the case where the Higgs mass is
twice the $W^\pm$ mass.  Compared to the natural scales of the system,
this solution has fairly small field amplitudes, but because of its large
spatial extent it is very massive.  Such large, coherent objects
are well described by the classical analysis undertaken here.
Quantization of the small oscillations around the oscillon solution
would nonetheless be of interest, perhaps using methods similar to
those applied to $Q$-ball oscillons in \cite{qqball}.  

Forming oscillons would likely require large energies
available only in the early universe.  In this context, it would be
very desirable to incorporate fermion couplings, which have been
ignored here.  (Of course, lattice chiral fermions introduce
significant, but not insurmountable, technical complications.)
While one might expect the oscillon to be destabilized by decay to
light fermions, in the case of the photon coupling we have
seen that the analogous decay mechanism is highly suppressed.  A
slow fermion decay mode would be of particular interest in
baryogenesis, since it would provide a mechanism for fermions to be
produced out of equilibrium, as is necessary to avoid washout of
particle/antiparticle asymmetry.   Or, if the oscillon is extremely
long-lived, it could provide a dark matter candidate.  If such results
proved compelling, this analysis would suggest a preferred value
of the Higgs mass.

\paragraph{Acknowledgments}

It is a pleasure to thank E.\ Farhi, F.\ Ferrer, A.\ Guth, R.\ R.\ Rosales,
R.\ Stowell and T.\ Vachaspati for helpful discussions, suggestions and
comments, and P.\ Lubans, C.\ Rycroft, S.\ Sontum, and P.\ Weakleim for
Beowulf cluster technical assistance.  N.\ G. was supported by National
Science Foundation (NSF) grant PHY-0555338, by a Cottrell College Science
Award from Research Corporation, and by Middlebury College.

Computational work was carried out on the Hewlett-Packard (HP) Opteron
cluster at the California NanoSystems Institute (CNSI) High
Performance Computing Facility at the University of California, Santa
Barbara (UCSB), supported by CNSI Computer Facilities and HP; the
Hoodoos cluster at Middlebury College; and the Applied Mathematics
Computational Lab cluster at the Massachusetts Institute of
Technology.  Access to the CNSI system was made possible through
the UCSB Kavli Institute for Theoretical Physics Scholars Program,
which is supported by NSF grant PHY99-07949.

\end{document}